\documentclass[twocolumn,showpacs,preprintnumbers,amsmath,amssymb]{revtex4}
\usepackage{graphicx}
\usepackage{dcolumn}
\usepackage{bm}
\begin{document}
\title{Temperature dependence of the resistance of metallic nanowires (diameter $\geq$ 15~nm): Applicability of  Bloch-Gr\"{u}neisen theorem.} 
\author{Aveek Bid\footnote[1]{Electronic mail: avik@physics.iisc.ernet.in}$^1$, Achyut Bora$^1$ and A. K. Raychaudhuri\footnote[2]{Electronic mail: arup@bose.res.in}$^{1,2}$}        
\address{$^1$Department of Physics, Indian Institute of Science,  Bangalore 560 012,  India\\
$^2$Unit for Nanoscience and Technology, S.N.Bose National Centre for Basic Sciences, Salt Lake, Kolkata 700098, India}
 
\begin{abstract}
We have measured the resistances (and resistivities) of Ag and Cu nanowires of diameters ranging from 15~nm to 200~nm in the temperature range 4.2~K-300~K with the specific aim to assess the applicability of the Bloch-Gr\"{u}neisen formula for electron phonon resistivity in these nanowires. The wires were grown within polymeric templates by electrodeposition.  We find that in all the samples   the resistance reaches a residual value at T=4.2~K and the temperature dependence of resistance  can be fitted to the Bloch-Gr\"{u}neisen formula in the entire temperature range with a well defined transport Debye temperature ($\Theta_{R}$). The  value of Debye temperature obtained from the fits lie within $8\%$ of the bulk value for Ag wires of diameter 15~nm while for Cu nanowires of the same diameter the Debye temperature  is significantly lesser than the bulk value.  The electron-phonon coupling constants (measured by $\alpha_{el-ph}$ or $\alpha_{R}$) in the nanowires  were found to have the same value as that of  the bulk.  The resistivities of the wires were seen to increase as the wire diameter was decreased. This increase in the resistivity of the wires may be attributed to surface scattering of conduction electrons. The specularity $p$ was estimated to be about 0.5. The observed results allow us to obtain the resistivities exactly from the resistance and  gives us a method of obtaining the exact numbers of wires within the measured array (grown within the template).
\end{abstract}

\maketitle
\section{Introduction}
Resistivity ($\rho$) of a metallic nanowire is a topic of considerable current interest. For nanowires with diameters approaching molecular dimensions the transport is likely to be quantum in nature~\cite{imry}. However,there is a considerable size range (diameter $>$ few nm) where the issues of quantum transport (like quantized conductance) are not important. Yet the study of the resistivity of such nanowire is of  interest because this is likely to be the dimension of metallic interconnects in electronic devices in the near future. In this size regime the concepts of Boltzmann transport are at its limits of applicability. A proper understanding of the resistivity in this regime is needed because it allows one to get a quantitative estimate of the resistance of the wire from its dimension without actually measuring it. In the regime where the width of the wire is a few tens of nm or less, it has been established adequately that $\rho$ is not determined by the material alone but by its size as well~\cite{welland,surface1, engelhardt, wuapl, silver}. For wires of these dimensions, the mean free path is comparable to or even larger than the wire width (particularly for clean wires) and one would expect the size effect to be operative. In this range, $\rho$  typically increases as the width of the wire is decreased. This is a serious issue in interconnects as an increase in the resistance of the wire increases the propagation delay time constant of the system and hence slows down the speed of the device. Recent investigations have focused on understanding the size effect in wires of width $<$ 100~nm so that a predictive relation can be obtained~\cite{welland, surface1, engelhardt, wuapl, silver}. The problem gets complicated due to the added contribution of grain boundary scattering. The size effect (arising mainly from the surface scattering) and the internal grain boundary scattering  (along with scattering form impurities or point defects) constitute the temperature independent part of the resistivity and this shows up as the residual resistivity at $T \leq 20K$ in most metallic solids. (If the wire is disordered the resistivity instead of showing a constant residual value at the low temperature can show an upturn often associated with effects such as localization ~\cite{bergmann}.)  Understanding these effects for a nanowire will thus give us a control on the residual resistivity of the nanowire.  However, understanding the residual resistivity alone is not enough to get a complete understanding of the $\rho$ in nanowires because at room temperatures, for a good metallic nanowire, a substantial part of the resistivity should arise from the temperature dependent part which in a non-magnetic metal arises from the electron-phonon interaction~\cite{ziman}. A good estimate and a quantitative way of predicting the resistivity contribution from the  electron-phonon interaction is thus needed. In this paper the principal focus is on the specific issue of the contribution of electron-phonon scattering to the resistivity of metallic nanowires and we establish by experiments to what extent such established theory as the Bloch-Gr\"{u}neisen  theory~\cite{ziman} is applicable as the wire diameter  goes down to as small as 15~nm.

Temperature dependence of the resistivity of nanowires have been studied both theoretically as well as experimentally ~\cite{wdw,dn,jph,ww,mt,rl}. Previous experimental studies on nanowires of elemental metals and alloys have established that in wires with diameter (or width) $\geq$ 15~nm, the temperature dependence of resistance is metallic reaching a residual resistivity at low temperature (if the wires are not disordered~\cite {wdw}) with $\rho \propto$ T for T$\geq$ 100~K. For wires with smaller diameter and width (typically $\leq$ 10~nm) the resistivity can have a negative temperature coefficient. This has been seen in wires of Au~\cite {wdw}, AuPd alloy~\cite{dn}, Zn~\cite{jph}, Cu~\cite{ww} and Sn~\cite{mt} (before it becomes superconducting at $T_{c}\approx$ 3.7~K). However, there has not been any experimental study that specifically addresses this issue in well characterized  nanowires over an extensive range of temperatures and wire diameters and analyzes the data quantitatively. With these objectives we have studied the resistance of metallic nanowires (silver and copper) as a function of the wire diameter (15~nm-200~nm) in the temperature range 4.2~K-300~K. The systematic investigation allows us to separate out the temperature independent part $R(0)$ from the temperature dependent part $\delta R(T)$ (we write $R(T)=R(0)+\delta R(T)$).  We analyzed the data in the framework of the Bloch-Gr\"{u}neisen  theory. This allowed us to estimate the effective Debye temperature from the temperature dependence of the resistance. In addition, a clear  evaluation of the residual resistivity allowed us to establish the dependence of the resistivity on the diameter of the wire. To our knowledge this is the first experimental study of the resistance of metallic nanowires over such an extensive range of temperature and size where the applicability of the Bloch-Gr\"{u}neisen formula has been tested quantitatively. 

\section{Electron–Phonon interaction and the Bloch-Gr\"{u}neisen formula}
The electron-phonon interaction and the Bloch-Gr\"{u}neisen formula have been adequately discussed in standard text books ~\cite{ziman}. However, for the purpose of quick reference as well as for completeness we give the important relations briefly in this section. In a non-magnetic metallic crystalline solid  the temperature dependence of the electrical resistivity arises mainly from the electron-phonon interaction ~\cite{ziman} and can be explained in the framework of the Boltzmann transport theory using the  Bloch-Gr\"{u}neisen formula:
\begin{equation*}
\rho(T)=\rho(0)+\rho_{el-ph}(T)
\end{equation*}
\begin{equation}
\rho_{el-ph}(T)=\alpha_{el-ph}( \frac{T}{\Theta_R})^n\int_0^{\frac{\Theta_R}{T}}\frac{x^n}{(e^x-1)(1-e^{-x})}dx
\label{bloch}
\end{equation}
where $\rho(0)$ is the residual resistivity due to defect scattering and is essentially temperature independent. The temperature dependent part of the resistivity $\rho_{el-ph}(T)$ arises from electron-phonon interaction. The constant $n$ generally takes the values 2,3 and 5 depending on the nature of  interaction and for a non magnetic elemental metal like Cu, Ag or Au with reasonable mean free path $n =5$.  $\alpha_{el-ph}$  is a constant that is $\propto \lambda_{tr}\omega_D /\omega_p^2$ where $\lambda_{tr}$ is the electron-phonon coupling constant, $\omega_D$ is the Debye frequency and $\omega_p$ is the plasma frequency ~\cite{allen}.  $\Theta_R$ is the Debye  temperature as obtained from resistivity measurements and matches very closely with the values of Debye temperature obtained from specific heat measurements. In the case of bulk silver and copper  with n=5 $\Theta_R \sim$ 200~K  and $\sim$ 320~K respectively ~\cite{mcdonald}. 

In Bloch-Gr\"{u}neisen formula, the phonons that contribute to the electron-phonon interaction are the acoustic phonons and one can get a simple one parameter scaling of the temperature dependence of the resistivity ($\rho$) where the only relevant parameter is the Debye temperature $\Theta_R$. For the specific case of $n=5$,
\begin{equation}
\frac {\rho_{el-ph}(T)}{\rho_{\Theta _R}}= \alpha_{R}( \frac{T}{\Theta_R})^5\int_0^{\frac{\Theta_R}{T}}\frac{x^5}{(e^x-1)(1-e^{-x})}dx
\label{bloch_scaling_rho}
\end{equation}
where $\rho_{\Theta _R}$ is the resistivity at the temperature $T=\Theta _R$. Thus the temperature dependence of the electrical resistance of a metal provides a useful method to estimate its effective Debye temperature. For acoustic phonons the value of $\alpha_{R}$ is 4.225~\cite{ross}. Even when there is an uncertainty in the  physical dimensions (that introduces uncertainty in the determination of $\rho$ from $R$) the above relation can be utilized because one can write 
\begin{eqnarray}
\frac{R(T)-R_{4.2K}}{R_{\Theta R}}=\frac{\delta R(T)}{R_{\Theta R}}=\frac {\rho_{el-ph}(T)}{\rho_{\Theta _R}} 
\label{bloch_scaling}
\end{eqnarray}

The measurements carried out in this paper  also allow us to investigate how the electron-phonon interaction gets modified in  nanowires due to size-reduction and manifests itself in the temperature dependence of the  resistance.  The lower limit of the integral in equation~\ref{bloch} being zero is the result of a tacit assumption that we are dealing with bulk material and hence the system size does not  impose a boundary condition on the maximum allowed phonon wavelength.  This assumption may no longer  be valid for wires of very small diameters. Thus the lower limit in the integral will now have a finite non-zero value. We wanted to investigate at what diameters of the wire does the size really begin to affect the phonon spectrum (and hence the electrical resistivity) of the wire.

\section{Experimental}
The nanowires of Ag and Cu with  average diameters in the range of 15~nm-200~nm and length 6$\mu$m were electrochemically deposited using polycarbonate membranes as template from AgNO$_3$ and CuSO$_4$ respectively ~\cite{PRB,akr1}. Schematic arrangement of the growth set-up is given in figure~1(a). During the growth, one of the electrodes was attached to one side of the membrane while the other electrode was a micro-tip of radius of curvature $\simeq 100 \mu m$ fitted to  a micropositioner. This electrode can be placed  at a specific area  on the membrane and the growth can thus be localized. The wires grow by filling the pores from end to end and as soon as one or more wires complete the path from one electrode to the other the growth stops. The wires after growth can be  removed from the membrane by dissolving the polymer in dichloromethane. This is needed for the microscopic characterization of the wires as described below. The wires after growth were annealed at 375K for 24 hours in vacuum under a dc current of  1~mA. The post-deposition annealing is needed to stabilize the resistance of the wires. For all successive measurements the current through the sample was kept lower than 100~$\mu$A. 

The structural and crystallographic nature of the wires form an important part in the analysis of  the data. The wires used in this investigation are single crystalline in nature. This has been established by such techniques as X-ray diffraction (XRD), Field Emission Gun-Scanning Electron Microscope (FEG-SEM)  and High resolution transmission electron microscope (HRTEM). The TEM was done in a Tecnai  G$^2$ 30 machine operated at 300KV with a nominal magnification of $10^6$. 
\begin{figure}[!htp]
\begin{center}
\vspace{-2cm}
\includegraphics[width=8.5cm]{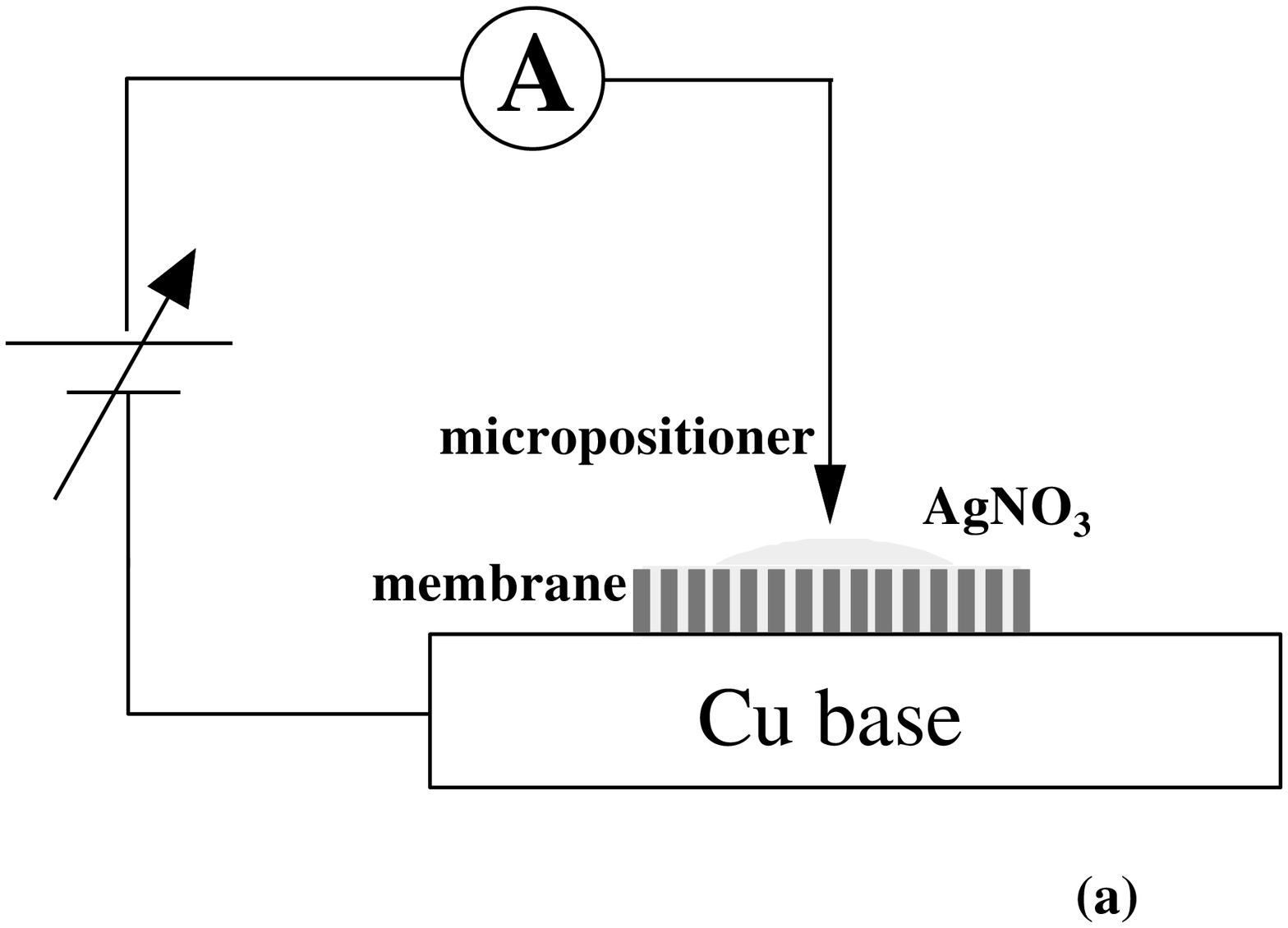}
\includegraphics[width=8.5cm]{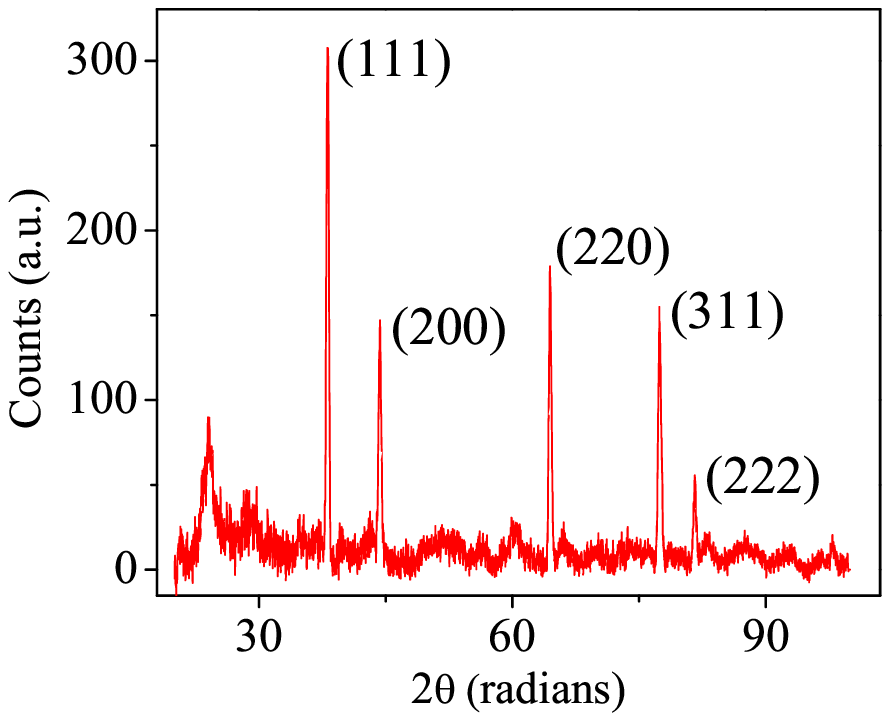}
\end{center}
\vspace{-1cm}
\caption{A schematic of the set up used for making the nanowires. (b) XRD of  100~nm Ag wires. The peaks have been indexed to FCC Ag.}
\end{figure}
\section{Basic results}
The XRD data are shown in  figure~1(b). The XRD data were indexed  as FCC lattice. The data as shown in figure~1(b) does not show any impurity peak. Similar data were also obtained for Cu and are not shown to avoid duplicity. 
\begin{figure}[!htp]
\begin{center}
\includegraphics[width=7cm]{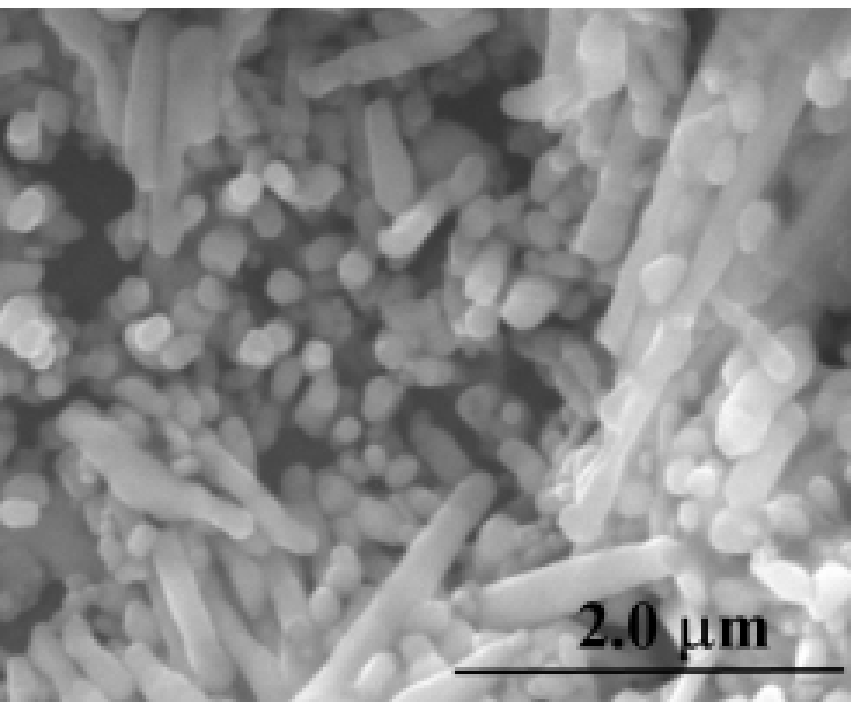}
\includegraphics[width=7cm]{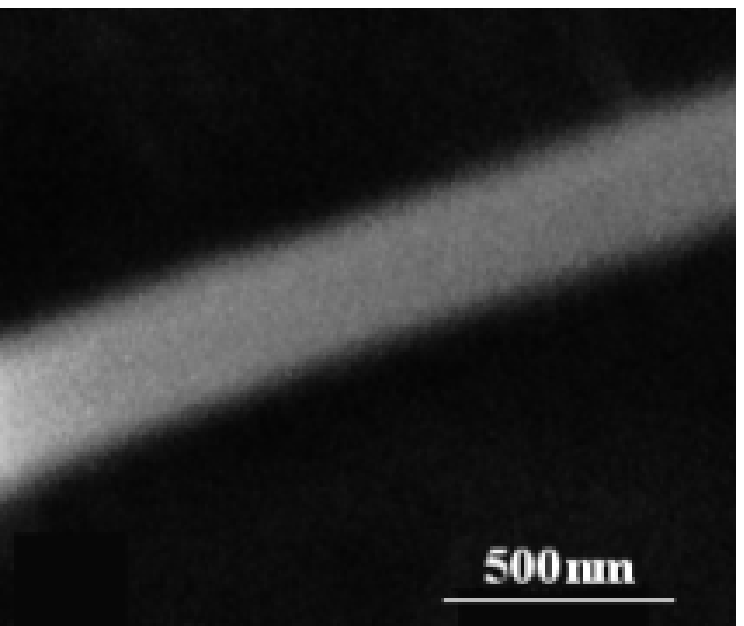}
\end{center}
\caption{SEM image of (a) 100~nm Ag wires taken after dissolving off the template in dichloromethane, (b) SEM image of a single 200~nm Ag wire.}
\end{figure}

Figure~2(a) shows the SEM image of a collection of  Ag  nanowires of diameter 100~nm taken after partially removing the membrane while figure~2(b) shows the SEM image of a single Ag wire of diameter 200~nm.  The average diameter of the wires match with the nominal diameter of the pores of the templates in which they were grown.
\begin{figure}[!htp]
\begin{center}
\includegraphics[width=7cm]{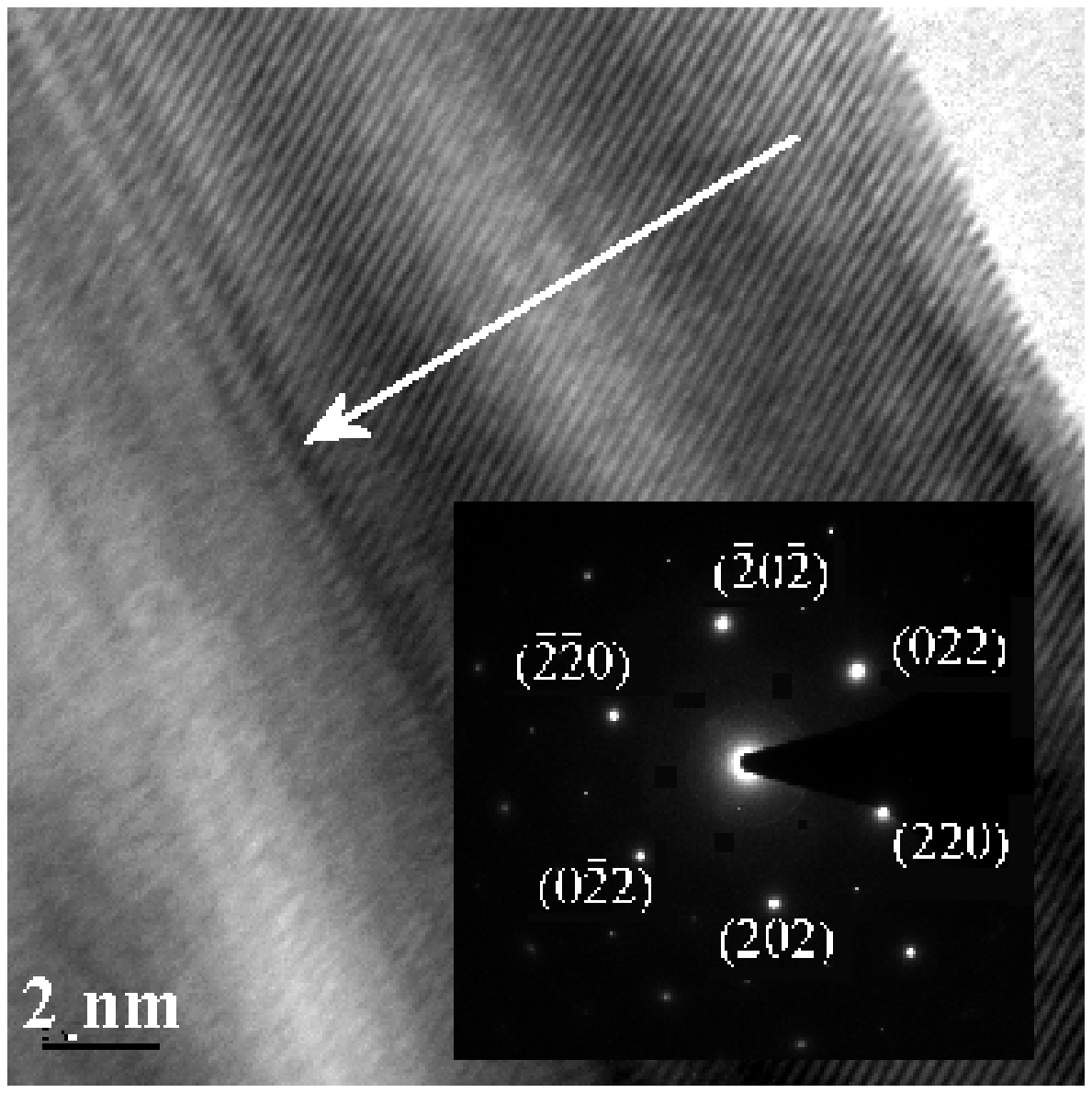}
\end{center}
\caption{HRTEM image of a 15~nm Ag wire. The arrow points the twin boundary present in the sample. The inset  shows the selective area diffraction pattern. The growth direction of the wires as seen from TEM is [100].}
\end{figure}

In figure~3 we show a typical TEM data taken on a 15~nm Ag wire.  The diffraction data are also shown and can be indexed into (220) planes.  The TEM data show that the wires are single crystalline with the presence of twins in them. They also show that there are no other substantial structural defects like grain boundaries or dislocations present in the  sample. The growth direction as seen from the TEM data is  [100].  

The resistance of the wires was measured in a bath type helium cryostat in the temperature range 4.2K-300K using an ac phase-sensitive detection using a lock-in amplifier. The measurements were carried out by retaining the wires within the polymeric membrane. On each of the two sides of the membrane two electrical leads were attached using silver epoxy.  Though the measurements were made with the wires retained within the membrane, the system is an array of parallel nanowires where the individual wires are well separated by the insulating membrane.  A typical sample may contain 2 to 50 wires. A very important issue in this measurement is the contribution of the contacts to the measurement.  We have paid attention to this aspect and measured the resistance  by making the contact in different ways.  In addition to silver epoxy contacts, we used  evaporated silver films to make contact which we find gave similar results. We also made contacts using wires tinned with Pb-Sn solder where we find that the change in the resistance of the sample as we go below the superconducting transition temperature ($\sim 7K$) of the solder is negligibly small ($\leq 2-3\%$).  All these tests rule out any predominant contribution from the contacts.  In the context of this paper we note that the small contact resistance even if it is present will make a small additive contribution to the temperature independent part of the resistance only. The particular issue of contact resistance as well as contact noise have been discussed in somewhat details in previous publications by the group~\cite{PRB,nanotech}. 
\begin{figure}[!htp]
\begin{center}
\includegraphics[width=8.5cm]{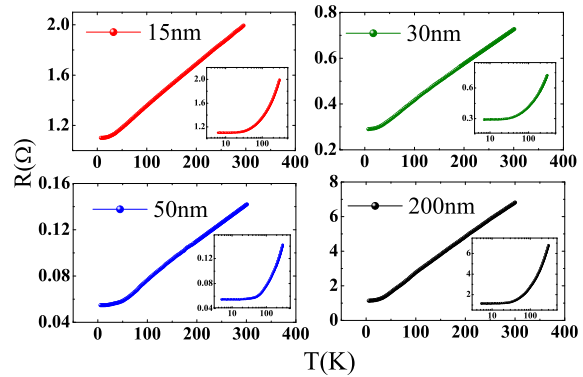}
\end{center}
\caption{Resistance of  arrays of Ag wires of diameter (a)15~nm, (b)30~nm, (c)50~nm and (d) 200~nm. The inset shows the resistance with the temperature axis in a  logarithmic scale in  order to show more clearly the low temperature behavior of the resistance.  }
\end{figure}
\begin{figure}[!htp]
\begin{center}
\includegraphics[width=8.5cm]{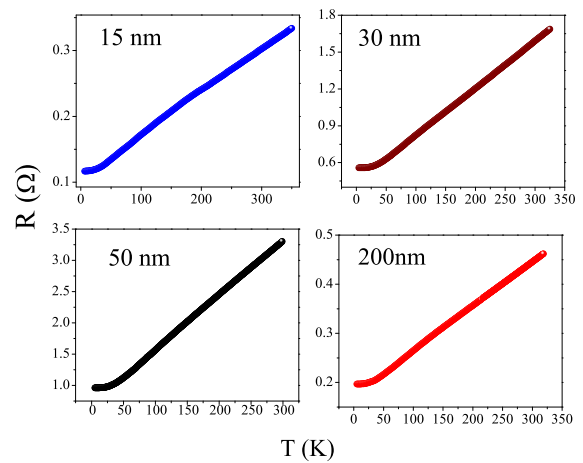}
\end{center}
\caption{Resistance of  arrays of Cu wires of diameter (a)15~nm, (b)30~nm, (c)50~nm and (d) 200~nm. }
\end{figure}

Figure~4 show the collection of resistance data on arrays of Ag wires with diameters ranging from 15~nm-200~nm. The insets show the same resistance plotted with a logarithmic scale for the temperature axis to show more clearly the low temperature behavior of the resistance. The data for the Cu nanowires are shown in figure~5.  The resistance data are typical of a good metal. Both the  Ag and Cu nanowire arrays have a fairly linear temperature dependence of resistance down to about 100~K  and reach a residual resistance below $40-50K$ with a  residual resistivity  ratio (RRR) $\frac{\rho_{300K}}{\rho_{4.2K}} \sim 3$.  The resistance  does not show any  upturn at low temperature thus ruling out any significant disorder in the system that can give rise to  effects such as localization ~\cite {tvr}. The value of  temperature coefficient of resistivity $\beta = 1/R(dR/dT)$ for the wires  lie within $\sim 4\times  10^{-3}$/K and $\sim 2.5\times  10^{-3}$/K respectively at 300~K. This matches well with the values for high purity Ag  and Cu  (approximately $3.8\times  10^{-3}$/K and $3.9\times  10^{-3}$/K respectively ~\cite{CRC}). This also emphasizes the essential defect free nature of the wires (as also established by the TEM images) as the presence of defects can reduce the value $\beta$ significantly. When the mean free path $l_{mfp}$ is so small that one reaches the Ioffe-Regel criterion $k_{F}l_{mfp} \approx 1$~\cite{ioffe}, $\beta$ becomes very small ($\ll 10^{-4}$)~\cite{mooij}. 

\section{Discussion}
\subsection{Applicability of Bloch-Gr\"{u}neisen theorem}
We fitted the observed resistance data to the Bloch-Gr\"{u}neisen function (equation ~\ref{bloch}) and  made the important observation that the resistance data for wires of all diameters (in the range 15~nm- 200~nm) could be fitted to the above mentioned  function  over the entire temperature range (4.2~K-300~K) of investigation  for integral values of $n$ using the Debye temperature ($\Theta_R$) and $\alpha_{el-ph}$  as the only two adjustable  fit  parameters. The fit parameters were optimized to give a  relative fit error (defined as ($R_{measured}-R_{fit})/R_{measured} \times 100 \%$)   of less than $\pm 0.5\%$ or better over the whole temperature range.  A typical fit is shown in figure~6 for Ag wires of diameter 15~nm.  In the inset we show the fit error which is less than $\pm 0.2\%$.
\begin{figure}[!htp]
\begin{center}
\includegraphics[width=8.5cm]{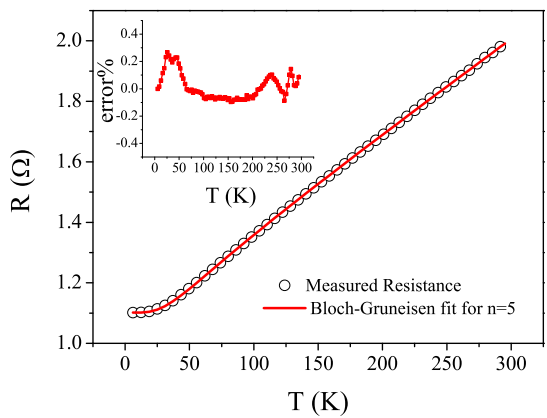}
\end{center}
\caption{Fit to Bloch-Gr\"{u}neisen formula (equation.~\ref{bloch}) to the measured resistivity data for 15~nm Ag wire. The inset shows the fit error (for definition see text) for $n=5$.}
\end{figure}
It was seen that  $n = 5$ gave the best fit for  the  wires.  The values of $\Theta_R$, as obtained from the fits (with $n=5$), are tabulated in table 1. Note that for the fit to the Bloch-Gr\"{u}neisen function we have used the resistance data directly because, as explained later, there is a substantial uncertainty in the determination of the number of wires in an array. The use of the resistance data for the fit does not change  $\Theta_R$. (Note:We use   $\alpha'_{el-ph}$ as a fit parameter in this case.This is related to $\alpha_{el-ph}$ by the relation  $\alpha'_{el-ph} =\frac{l}{NA}\alpha_{el-ph}$. $l$ and $A$  are the length and area of cross-section respectively and are known from experiment.Using the method discussed later in this paper (and appendix-A) we could calculate the value of $N$ more accurately and hence the value of  $\alpha_{el-ph}$ from the fit parameter $\alpha'_{el-ph}$. The value of $\alpha_{el-ph}$  thus estimated is $\approx 4.6\times10^{-8}$~$\Omega$m for  the Ag and  $\approx 4\times10^{-8}$~$\Omega$ m for Cu wires. This implies that the coupling constant is essentially unchanged on size reduction.
\begin{table}
\caption{\label{tab:table1}The values of $\Theta_R$ and $\alpha_R$  (see equation.~\ref{bloch_scaling})  obtained from resistance data for Ag and Cu wires of different diameters. $\Theta_R$ for bulk Ag and Cu are 200K and 320K respectively.}
\begin{ruledtabular}
\begin{tabular}{ccccc}
Sample&diameter (nm)&$\Theta_R (K)$&\% reduction&$\alpha_{R}$  \\
\hline
Ag&15 &184&8&4.226 \\
Ag&30 & 170&15&4.227\\
Ag&100 &174&13&4.225 \\
Ag&200&187&6.5&4.226\\
Cu&15&180&43&4.271\\
Cu&30&235&26&4.224\\
Cu&50&231&28&4.225\\
Cu&200&200&37&4.235\\
\end{tabular}
\end{ruledtabular}
\end{table}

The  values of $\Theta_R$ for all the samples measured were found to be close  to but significantly smaller than the bulk value as measured in a reference sample. (For a reference sample we used a thermally evaporated  high quality Ag thin film of thickness 150~nm. The film was evaporated in UHV chamber with a base pressure $10^{-8}$ mbar from an effusion cell. It has a room temperature resistivity of $1.63 \times 10^{-8}$~$\Omega$ m and a residual resistivity ratio of about 10. $\Theta_R$ for this sample matches with that of the bulk.)  Analysis of the data as summarized in table 1, indicate that $\Theta_R$ has a reduction of $\approx 8\%-15\%$ for Ag and $\approx 25\%-40\%$ for Cu. The analysis thus gives us a direct way to estimate the Debye temperature in the nanowires. The observation of a softening of $\Theta_R$ is significant as also the fact that it is material dependent, being strong in Cu and not so significant in Ag.

\begin{figure}[!htp]
\begin{center}
\includegraphics[width=8.5cm]{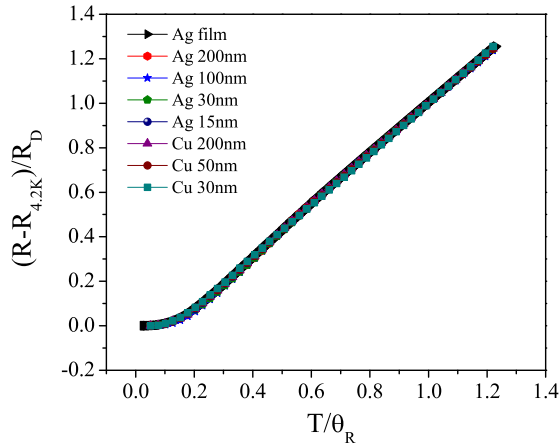}
\end{center}
\caption{Plot of   $\frac{R_{T}-R_{4.2K}}{R_{D}}$ as a function of $\frac{T}{\Theta_R}$  (see equation.~\ref{bloch_scaling}).  Here $R_D$ is the value of the measured resistance at T=$\theta_R$.}
\end{figure}
The applicability of the Bloch-Gr\"{u}neisen theorem can be better appreciated when we analyze the data using the scaling equation ~\ref{bloch_scaling_rho}.  We checked the  scaling equation by plotting  (see figure~7)  $\frac{R_{T}-R_{4.2K}}{R_{D}}$ as a function of $\frac{T}{\Theta_R}$.  Here $R_D$ is the value of the measured resistance at T=$\theta_R$. It can be seen that all the resistance data collapse into one curve signifying that the one parameter scaling law  holds. From table I it can be seen that the value of the constant $\alpha_{R}$  is roughly the same for all the nanowires and is the value (=4.225) predicted by the simple acoustic phonon-electron coupling theory ~\cite{ross}.  The observation that the Bloch-Gr\"{u}nesien theorem is quantitatively applicable in the nanowires of at least two materials studied down to 15~nm diameter is extremely important. It establishes that the  temperature dependent part of the resistivity (as arising from the electron-acoustic phonon interaction) is unchanged on size reduction down to 15~nm. The implication of the observation  is that one now has a tool with which one can obtain the electron-phonon contribution to the resistivity with good quantitative accuracy. 
We will show below that this basic observation can be utilized  to calculate/estimate the resistivity of metallic nanowires even if the exact number of wires in an array is not known.

The number of wires in such a sample of array vary from about 2 to 50 as was estimated using the bulk value of the resistivity. This is a very rough estimate as the resistivity in this range depends on the diameter/width of the wire. Hence we do not use this in our calculations. It is to be noted that the exact value of the resistivity is not needed to check the applicability of the Bloch-Gr\"{u}neisen theorem as the analysis for the determination of $\Theta_R$ were carried out using the resistance instead of the resistivity. Neither is the exact value of  $\rho$ needed to check equation ~\ref{bloch_scaling_rho}. As a result the exact number of wires in the array was not needed for any of the evaluation so far.  Below we describe a method of estimating the resistivity of the samples which also gave us a better estimate of the number of wires in each sample. 

\subsection{Dependence of phonon contribution to resistance on the size of the wire}
The applicability of the Bloch-Gr\"{u}neisen theorem would imply that the basic electron-acoustic phonon interaction as well as the simple Debye phonon spectrum (phonon density of states  $\propto \omega^{2}$ where $\omega$ is the phonon frequency ) remains unchanged on size reduction. This would raise the question - to what size one can reduce the wire diameter before the deviation from simple Bloch-Gr\"{u}neisen theorem begins to show up. To estimate the effect of  size reduction, if any, on the phonon contribution to the resistance of a wire, we study the variation of the  integrand of  equation ~\ref{bloch} as a function of $x$ (where $x$ = $\Theta/T$ = $hc/\lambda k_BT$. Here $c$ is the sound velocity averaged over all the acoustic modes and $\lambda$ is the phonon wavelength.) At a given temperature T the dominant contribution to the integral in equation~\ref{bloch} comes from the value of x for which the integrand has a maximum. We define this  value of $x$ as $x_{d}$, the dominant value of $x$ at a given temperature.  $x_{d}$ depends on the value of n ( for $n=5$, $x_{d} = 5$). Thus at a temperature T the phonons having a wavelength $\approx$ $\lambda_{d}=hc/x_{d}k_BT=hc/5k_BT$ are the dominant phonons that contribute most to the temperature dependent part of the resistance of a metal. The values of $\lambda_{d}$ (for $n=5$) as a  function of temperature  T are plotted in figure~8. 
\begin{figure}[!htp]
\begin{center}
\includegraphics[width=8.5cm]{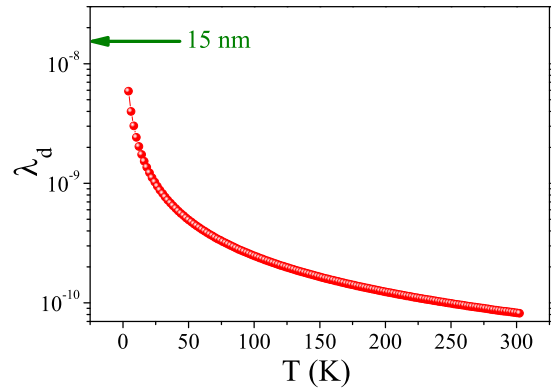}
\end{center}
\caption{Plot of the dominant phonon wavelength $\lambda_d$ as a function of temperature for $n=5$. The arrow points to 15~nm, the diameter of the narrowest wire measured by us.}
\end{figure}
We can see from the plot that even at 4.2~K, the value of $\lambda_d$ is much smaller than 10~nm. Thus for a wire of diameter 15~nm, the phonons that make the maximum contribution to the temperature dependent part of the resistance (down to the lowest temperature measured) are not affected by the wire dimensions. At higher temperatures $\lambda_d$ becomes smaller ( being $\propto 1/T$). So in all the samples studied by us down to the  lowest diameter  phonon confinement is not an issue and hence contribution to the temperature dependent part of the resistivity is expected to follow the same Bloch-Gr\"{u}neisen function valid for a 3D bulk metal albeit with a reduced $\Theta_R$.  To test the effect of phonon confinement and to evaluate the effect of finite size of the wire on the electron-phonon part of the resistivity, we have fitted the resistance data to the Bloch-Gr\"{u}neisen function with the lower limit of the integral changed to $x_{min}=\theta_{min}/T$ where $\theta_{min} = hc/dk_B$ ($d$ being the diameter of the wire).  We do not find any significant change in the quality of the fit or  in the value of the fit parameters. This is because even for wires of diameter 15~nm, $x_{min} \sim 0.5$ at 30K and the values of $x$ in this range hardly make any contribution to the Bloch-Gr\"{u}neisen integral in equation ~\ref{bloch}. Presumably at even lower temperatures $T < 0.5K$ one would expect that for wires of this size the finite size effect should show-up in the Bloch-Gr\"{u}neisen estimate of the temperature dependent part of the resistivity $\rho$. However, in this temperature range in a real wire the residual resistivity will mask any effect of the temperature dependent part.

\subsection{Estimation of resistivity from the resistance data and its dependence on wire diameter}
We noted before that the  evaluation of the resistivity from the directly observed resistance data  is prone to error  due to the uncertainty in the actual number of wires that make up the array. The observation that the phonon contribution  to the electrical resistance of the nanowires is unchanged from that of the bulk allows us to estimate the value of the resistivity of the wires from the measured resistance without having to know the number of wires present in the sample. We elaborate this in Appendix-A. Let the resistance of one wire be  $R_1 = \rho l/A$, where $\rho$ is the resistivity of the material, $l$ is the length and $A$ is the cross-sectional area of the wire.  Therefore resistance of N identical wires in parallel is 
\begin{equation}
R_N =  \rho l/AN. 
\label{rn}
\end{equation}

The facts that the temperature dependence of the resistivity arises from the electron-phonon interaction and that  the same relation (with the same $\alpha_{el-ph}$ and $\alpha_R$) governs the electron-phonon interaction in the nanowire as in the bulk crystalline material can be utilized to get  (see appendix-A):
\begin{equation}
\rho=\rho_0\frac{\beta_0}{\beta}\frac{\Theta_{R0}}{\Theta_R}
\label{rn1}
\end{equation}
where $\beta$   $=[1/R_N(dR_N/dT)]$ is the measured temperature coefficient of resistivity (TCR) of the wire array, $\beta_0$ is the temperature coefficient of resistivity of the bulk material, $\rho_0$ is the bulk resistivity at that temperature and $\Theta_{R0}$ is the Debye temperature of the bulk material.  (This equation is valid strictly in the high temperature limit where the temperature dependent part of the electrical resistivity of a metal  is only due to   phonon scattering. In the nanowires used by us the temperature dependent part solely arises from the electron-phonon interaction, as established by analysis of the data in the previous section.)

We calculate the  TCR from the measured resistance near room temperature.  Using the resistivity at 295K thus estimated and from the measured  resistance of the wire arrays we can find the number of wires in a given array (N) using equation ~\ref{rn}. From this we can evaluate the resistivity as a function of temperature for all the wires. This is shown  in figure~9 for Ag wires as well as the Cu wires. 
\begin{figure}[!htp]
\begin{center}
\includegraphics[width=8.5cm]{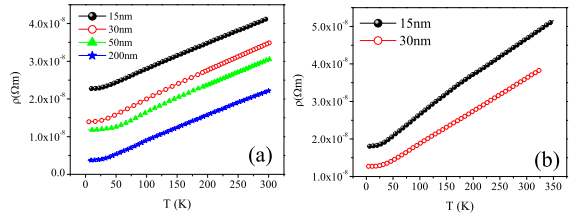}
\end{center}
\caption{Plot of the resistivity of Ag and Cu  nanowires  as a function of temperature. The resistivity were obtained using the method described in the text.}
\end{figure}
One can see a  systematic variation of $\rho$ as a function of temperature for wires of different diameters. The value of the resistivity at 295K, the residual resistivity as well as the mean-free path ($l_{mfp})$ of the electrons calculated from the resistivity at T=295K are shown in table 2. For this calculation we assumed bulk electron density. In figure~10(a) we also plot the residual resistivity $\rho_{4.2K}$ as well as the $\rho_{295K}$  as a function of $d$ for the Ag wires. (The bulk value of the resistivity of high purity Ag  ($1.6 \times 10^{-8} \Omega m$) at 295K  is shown as an arrow.)  We have plotted the ratio  $l_{mfp}/d$ at 295~K as a function of $d$  in figure~10(b).  It clearly shows that in wires of small diameters the $l_{mfp}$ can be substantially larger than $d$ and that the ratio $l_{mfp}/d$ increases as $d$ is reduced. For wires of diameter 15~nm  $l_{mfp}/d$ approaches a value 1.5. We also note that the value of $k_Fl$ is significantly larger than 1 thus justifying the applicability of the Bloch-Gr\"{u}neisen theorem in these samples.
\begin{figure}[!htp]
\begin{center}
\includegraphics[width=8.5cm]{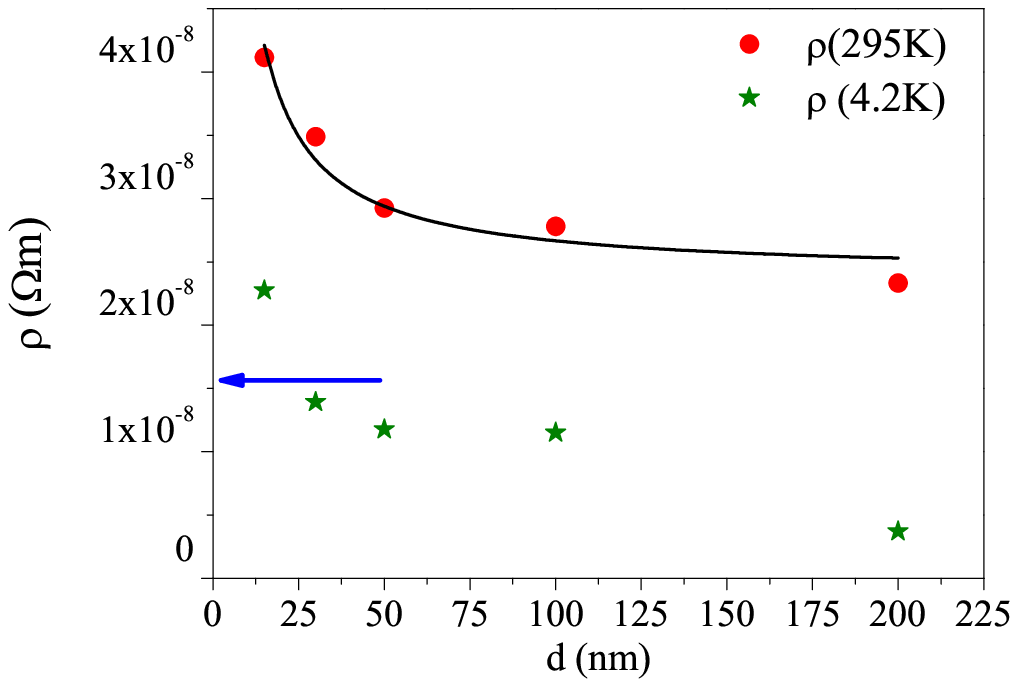}
\includegraphics[width=8.5cm]{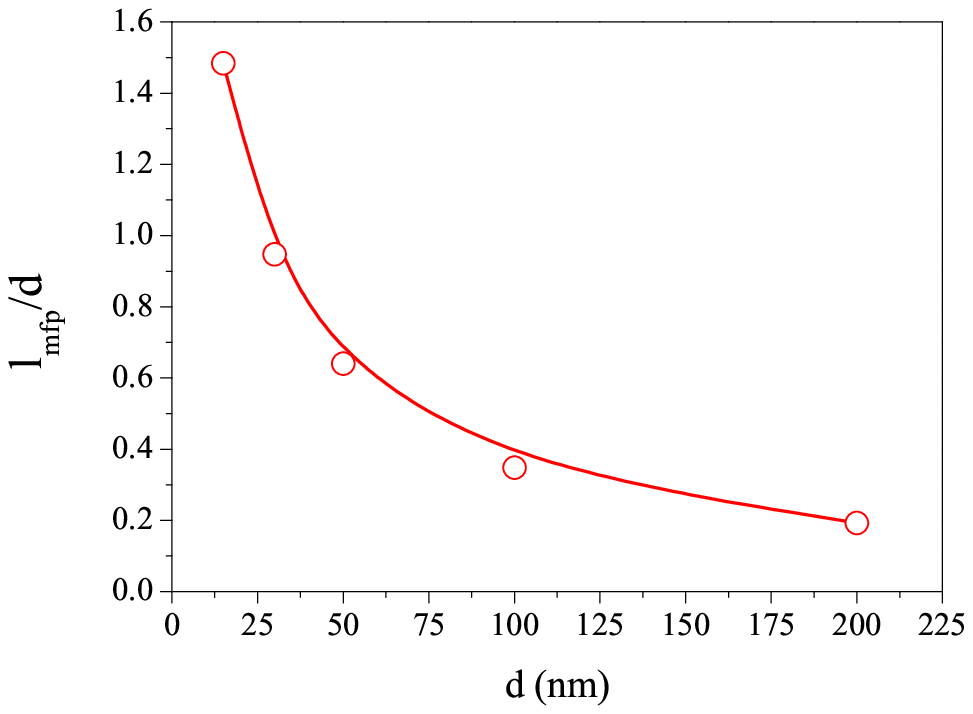}
\end{center}
\caption{(a) Plot of the resistivity of Ag nanowires at 295K and the residual resistivity as a function of the wire diameter. The line is the fit to the data using equation~\ref{dingle}. The arrow is the bulk value of the resistivity of Ag $1.629 \times 10^{-8} \Omega m$. (b) Plot of  $l_{mfp}/d$ as a function of $d$ for Ag nanowires. The solid line is a guide to the eye. }
\end{figure}

\begin{table}
\caption{\label{tab:table2}The values of $\rho$ at  295~K, the residual resistivity at 4.2~K  and $l_{mfp}$ at 295~K  for wires of different diameters.}
\begin{ruledtabular}
\begin{tabular}{ccccc}
Sample&diameter&$\rho_{295K}$&$\rho_{4.2K}$&$l_{mfp}$\\
&(nm)&($\mu\Omega$cm)&($\mu\Omega$cm)&(nm)\\
\hline
Ag& 15&4.11&2.27&20 \\
Ag& 30&3.49&1.39&24\\
Ag& 50&2.92&1.17&29\\
Ag&100&2.78&1.15&30\\
Ag&200&2.33&0.37&36\\
Cu&15&5.67&2.2&12\\
Cu&30&3.58&1.27&18\\
Cu&50&3.09&1.02&21\\
\end{tabular}
\end{ruledtabular}
\end{table}

From table~\ref{tab:table2} as well as from figures~9 and 10 we can see  that there is a significant increase in the  residual resistivity as the diameter of the wires is decreased.  The increase is systematic. All wires have the same chemical purity and hence the  enhancement of $\rho_{4.2K}$ on reduction of $d$ is not due to impurity contribution. For wires in the diameter range that we are studying, the resistivity is expected to increase with a decrease in the wire diameter due to 2 probable reasons:
\begin{itemize}
\item{As the mean free path of the electrons is now of the order (and in some cases larger than) of the diameter of the wire, the bulk assumption is no longer valid and the electron starts seeing the external surface of the sample. The scattering of the electrons from the boundaries of the sample increases the resistivity.  This effect is generally quantified by the specularity parameter  $p$ that represents the fraction of  conduction electrons reflecting specularly from the surface~\cite{fs}.}
\item{As the mean grain size in narrow polycrystalline wires/thin films is generally of the same order as  the diameter/thickness, a reduction in the sample size increases the number of grain boundaries. This excess grain boundary scattering  leads to an increase in the resistivity of the material over that the bulk value~\cite{ms}.}
\end{itemize}
 The TEM images show that  there are  no significant presence of grain boundaries in the wires studied by us. The twin boundaries seen are generally weak scatterer and do not contribute  significantly to the resistivity.  We analyze the resistivity data in the light of  surface scattering model of Dingle~\cite{dingles} and Chambers~\cite{chambers}. According to these authors, for cylindrical wires of isotropic metal, the dependence of the resistivity  on the wire diameter goes as:
\begin{equation}
\rho_d=\rho_0+\rho_{0}(1-p)l_{0}/d
\label{dingle}
\end{equation}
where $\rho_d$ is the resistivity of  the wire of diameter $d$, $\rho_{0}$ is the $\lq\lq$bulk" or the so called $\lq\lq$intrinsic" resistivity for the material of the wire, $p$ is the specularity that denotes the fraction of conduction electrons  undergoing specular reflection at the wire surface and  $l_{0}$ is the bulk mean free path. The quantity $\rho_{0} l_{0}$ is the property of the wire and is temperature independent. To check the validity of the model we have evaluated the value of  $\rho_{0} l_{0}$ at several different temperatures and got similar values to within $\pm 3\%$.  For Ag wires $(\rho_{0} l_{0} \approx 5.5 \times 10^{-16} \Omega m^2$.(Note: In this analysis we use only the data of Ag wires as they are of higher chemical purity and contain less density of defects.)  In figure 10(a) we show calculated value of  $\rho_d$ for $p \approx 0.5$. The calculated values  closely match the observed data. The value of specularity factor $p \approx 0.5$ obtained from our data  is very close to the values (0.3-0.5) reported previously for metallic films of similar dimensions ~\cite{welland, engelhardt}.  

To summarize, we have measured the resistances (and resistivities) of Ag and Cu nanowires of diameters ranging from 15~nm upto 200~nm in the temperature range 4.2-300K. We find that the temperature dependence of resistance  can be fitted to a Bloch-Gr\"{u}neisen formula in the entire temperature range. This ensures that the Debye temperature is a viable parameter and this allows us to obtain a value for the Debye temperatures.The  values of Debye temperature obtained form the lay the fits within $8\%$ of the bulk value for Ag wires of diameter 15~nm. However, there is significant softening of the Debye temperature for Cu nanowires with the same diameter. The electron-phonon coupling constants (measured by $\alpha_{el-ph}$ or $\alpha_{R}$) in the nanowires  were found to have the same value as the bulk.  The resistivities of the wires were seen to increase as the wire diameter was decreased. This increase in the resistivity of the wires may be attributed to surface scattering of conduction electrons. The specularity $p$ was estimated to be about 0.5. 

\section{Acknowledgments} 
This work is  supported by DST, Govt. of India and CSIR, Govt. of India. The first author (A.B.) acknowledges CSIR for support and the second author (A.B.) acknowledges UGC for support.
\appendix
\section{}
The resistivity $\rho$ can be written as  $\rho(T)=\rho(0)+\rho_{el-ph}(T)$ where  $\rho_{el-ph}(T)$  is the temperature dependent part of the resistivity due to the phonons and  $\rho(0)$  is the residual resistivity due to defect and surface scattering. The fact that the grain boundary scattering and surface scattering contributions to the resistivity is independent of  temperature (at least at high temperatures) has recently been shown experimentally~\cite{surface1}.  So we can have 
\begin{equation}
\frac{d\rho}{dT}=\frac{d\rho_{el-ph}}{dT}
\end{equation}
At high temperatures, $d\rho/dT \propto 1/\Theta_R$ and since the coupling constant $\alpha_{el-ph}$ is nearly the same for all the nanowires with the same chemical elements (established by this experiment) we can write:
\begin{equation}
\frac{d\rho_{el-ph}}{dT}=\frac{d\rho_{(el-ph)0}}{dT}\frac{\Theta_{R0}}{\Theta_R}=\frac{d\rho_0}{dT}(\frac{\Theta_{R0}}{\Theta_R})
\end{equation}
where $\rho_0$ refers to the bulk resistivity of the material and $\Theta_{R0}$ is the bulk Debye temperature. From equation ~\ref{rn} we get:
\begin{eqnarray}
\frac{dR_N}{dT}=\frac{l}{NA}\frac{d\rho}{dT}=\frac{l}{NA}\frac{d\rho_0}{dT}\frac{\Theta_{R0}}{\Theta_R}
\nonumber\\
=\frac{l \rho_0}{NA}\frac{\Theta_{R0}}{\Theta_R}\beta_0=\frac{R_N\rho_0}{\rho}\frac{\Theta_{R0}}{\Theta_R}
\beta_0
\label{tcr}
\end{eqnarray}
where $\beta_0$ is the temperature coefficient of resistivity of the bulk material. Equation~\ref{tcr} immediately yields 
\begin{equation}
\rho=\rho_0\frac{\beta_0}{\beta}\frac{\Theta_{R0}}{\Theta_R}
\label{rho}
\end{equation}
where $\beta$  $=[1/R_N(dR_N/dT)]$ is the measured temperature coefficient of resistivity of the wire array.

\end{document}